\documentclass[fleqn,usenatbib]{mnras}

\usepackage{newtxtext,newtxmath}
\usepackage{amsmath}
\usepackage{courier}
\usepackage{threeparttable}
\usepackage{float}
\usepackage{xcolor}
\usepackage{blindtext}
\usepackage{scrextend}
\usepackage{colortbl}

\usepackage[T1]{fontenc}
\usepackage{ae,aecompl}

\usepackage{graphicx}	   
\usepackage{amsmath}	   
\usepackage{amssymb}	   
\usepackage{physics}       
\usepackage{algorithm}     
\usepackage{algpseudocode}

\title[Sparse RFI prediction using FCNs]{Optimizing Sparse RFI Prediction using Deep Learning}

\author[J. R. Kerrigan et al.]{
\large
Joshua Kerrigan$^{1}$\thanks{E-mail: joshua\_kerrigan@brown.edu (JRK)},
Paul La Plante$^{2}$,
Saul Kohn$^{2}$,
Jonathan C. Pober$^{1}$,
James Aguirre$^{2}$,
Zara Abdurashidova$^{3}$,
\newauthor
\large
Paul Alexander$^{4}$,
Zaki S. Ali$^{3}$,
Yanga  Balfour$^{5}$,
Adam P. Beardsley$^{6}$,
Gianni  Bernardi$^{5,7,8}$,
Judd D. Bowman$^{6}$,
\newauthor
\large
Richard F. Bradley$^{9}$,
Jacob  Burba$^{1}$,
Chris L. Carilli$^{4,10}$,
Carina  Cheng$^{3}$,
David R. DeBoer$^{3}$,
Matt  Dexter$^{3}$,
\newauthor
\large
Eloy  de~Lera~Acedo$^{4}$,
Joshua S. Dillon$^{3}$,
Julia Estrada$^{19}$,
Aaron  Ewall-Wice$^{11}$,
Nicolas  Fagnoni$^{4}$,
Randall  Fritz$^{5}$,
\newauthor
\large
Steve R. Furlanetto$^{12}$,
Brian  Glendenning$^{10}$,
Bradley  Greig$^{13,20}$,
Jasper  Grobbelaar$^{5}$,
Deepthi Gorthi$^{3}$,
Ziyaad  Halday$^{5}$,
\newauthor
\large
Bryna J. Hazelton$^{14,15}$,
Jack  Hickish$^{3}$,
Daniel C. Jacobs$^{6}$,
Austin  Julius$^{5}$,
Nick  Kern$^{3}$,
Piyanat  Kittiwisit$^{6}$,
\newauthor
\large
Matthew  Kolopanis$^{6}$,
Adam  Lanman$^{1}$,
Telalo  Lekalake$^{5}$,
Adrian  Liu$^{16}$,
David  MacMahon$^{3}$,
Lourence  Malan$^{5}$,
\newauthor
\large
Cresshim  Malgas$^{5}$,
Matthys  Maree$^{5}$,
Zachary E. Martinot$^{2}$,
Eunice  Matsetela$^{5}$,
Andrei  Mesinger$^{17}$,
Mathakane  Molewa$^{5}$,
\newauthor
\large
Miguel F. Morales$^{14}$,
Tshegofalang  Mosiane$^{5}$,
Abraham R. Neben$^{11}$,
Aaron R. Parsons$^{3}$,
Nipanjana  Patra$^{3}$,
\newauthor
\large
Samantha  Pieterse$^{5}$,
Nima  Razavi-Ghods$^{4}$,
Jon  Ringuette$^{14}$,
James  Robnett$^{10}$,
Kathryn  Rosie$^{5}$,
Peter  Sims$^{1}$,
\newauthor
\large
Craig  Smith$^{5}$,
Angelo  Syce$^{5}$,
Nithyanandan  Thyagarajan$^{6,10}$,
Peter K.~G. Williams$^{18}$,
Haoxuan  Zheng$^{11}$
\vspace{0.4cm} \\
\footnotesize
The authors' affiliations are shown in Appendix~\ref{appendix:affil}
}
\date{Accepted XXX. Received YYY; in original form ZZZ}

\pubyear{2018}

\begin{document}
\label{firstpage}
\pagerange{\pageref{firstpage}--\pageref{lastpage}}
\maketitle

\begin{abstract}
Radio Frequency Interference (RFI) is an ever-present limiting factor among radio telescopes even in the most remote observing locations. When looking to retain the maximum amount of sensitivity and reduce contamination for Epoch of Reionization studies, the identification and removal of RFI is especially important. In addition to improved RFI identification, we must also take into account computational efficiency of the RFI-Identification algorithm as radio interferometer arrays such as the Hydrogen Epoch of Reionization Array grow larger in number of receivers. To address this, we present a Deep Fully Convolutional Neural Network (DFCN) that is comprehensive in its use of interferometric data, where both amplitude and phase information are used jointly for identifying RFI. We train the network using simulated HERA visibilities containing mock RFI, yielding a known ``ground truth'' dataset for evaluating the accuracy of various RFI algorithms. Evaluation of the DFCN model is performed on observations from the 67 dish build-out, HERA-67, and achieves a data throughput of 1.6$\times 10^{5}$ HERA time-ordered 1024 channeled visibilities per hour per GPU. We determine that relative to an amplitude only network including visibility phase adds important adjacent time-frequency context which increases discrimination between RFI and Non-RFI. The inclusion of phase when predicting achieves a Recall of 0.81, Precision of 0.58, and $F_{2}$ score of 0.75 as applied to our HERA-67 observations.
\end{abstract}




\section{Introduction}

Next generation radio interferometers are now beginning to become operational. These arrays are looking to detect and measure some of the weakest signals the Universe has to offer, such as the brightness-temperature contrast of the 21cm signal during the Epoch of Reionization (EoR). By measuring this highly redshifted signal we can characterize the progression of the EoR. The understanding gained from this characterization has the potential to help us unravel how the first stars and galaxies formed and reionized their surrounding neutral hydrogen. While instruments like the Hydrogen Epoch of Reionization Array (HERA) \citep{DeBoer2017} have the intrinsic sensitivity required to detect the EoR signal through a power spectrum, they are afflicted with anthropogenic noise which we refer to as Radio Frequency Interference (RFI). Interference from RFI in 21cm EoR observations is an especially significant obstacle because it can have a brightness anywhere from on the order of the EoR signal to orders of magnitude beyond even Galactic and extra-galactic foregrounds. RFI unfortunately introduces a reduction in sensitivity in two separate but distinct ways, one being the direct contamination by having similar spectral characteristics and overpowering of the 21cm signal, and the other being the introduction of a complex sampling function due to missing data. This produces correlations between modes when computing the Fourier transform along the frequency axis \citep{Offringa2019}. It is therefore important to strike a balance between identifying RFI while not falsely identifying non-RFI as RFI, which leads to further complicating our sampling function over frequency.
Many approaches have recently been developed to identify and extract RFI from radio telescope data. RFI algorithms of particular interest include AOflagger \citep{Offringa2012}, which uses a Scale-invariant Rank operator to identify morphologies that are scale-invariant in time or frequency which is a characteristic of many RFI signals. This RFI detection strategy has been used successfully on instruments such as the MWA \citep{Offringa2015} and the Low-Frequency Array (LOFAR) \citep{OffringaLOFAR2013}. Alternative approaches to RFI identification include the application of neural networks. More specifically, a Deep Fully Convolutional Neural Network (DFCN) based on the U-Net architecture \citep{Ronneberger2015} has been used on single dish radio telescope data \citep{Akeret2017}, and a Recurrent Neural Network (RNN) has been applied to signal amplitudes from radio interferometer data \citep{Burd2018}. 

\begin{figure}
\includegraphics[scale=2.5,width=\columnwidth]{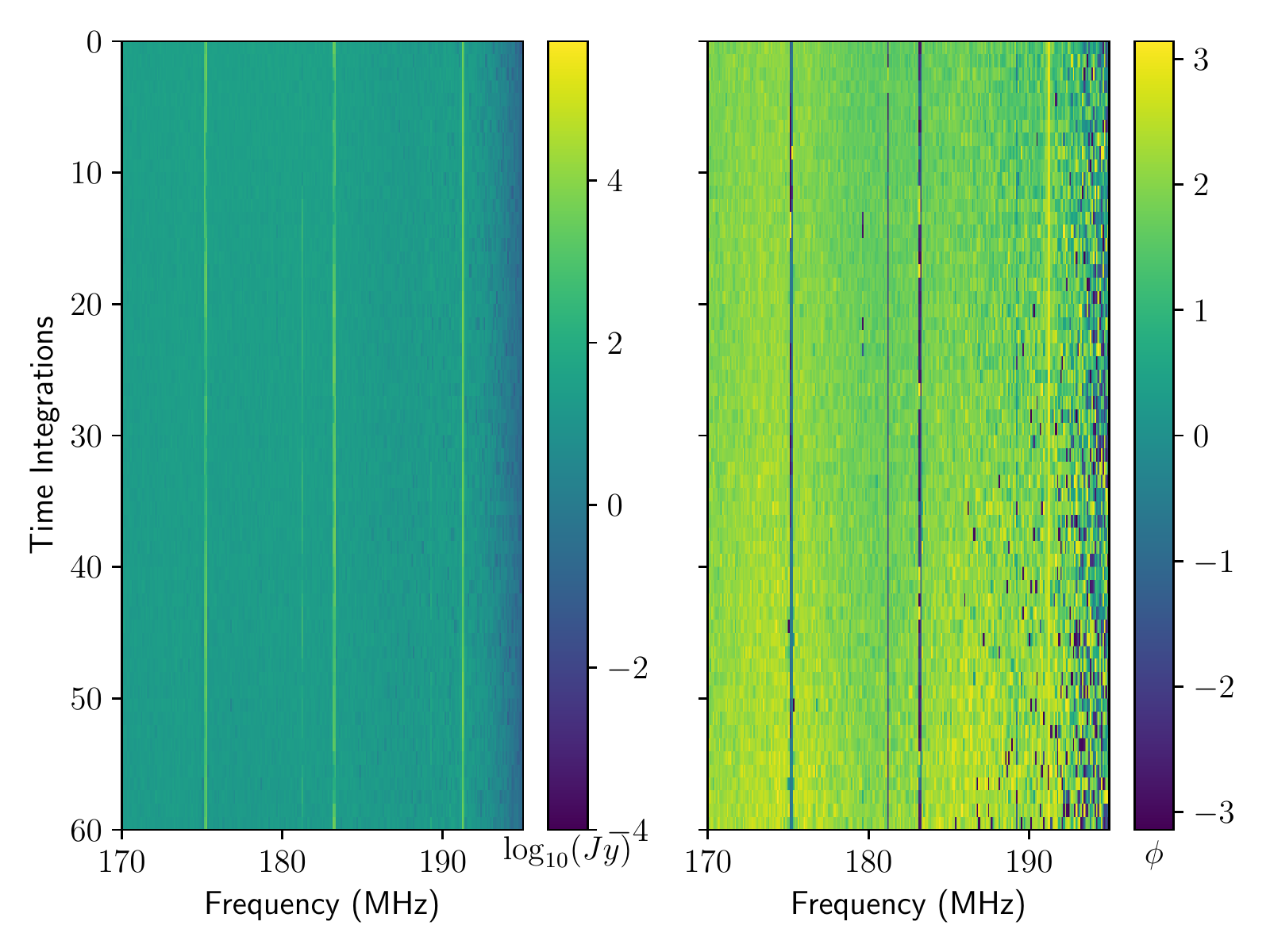}
\caption{An example of a HERA 14m baseline waterfall visibility between 170-195 MHz in amplitude (left) and phase (right). The phase waterfall visibility demonstrates how it can provide complementary information about the presence of RFI such as in the 181.3 MHz channel which has constant narrow-band RFI and the more spontaneous `blips' in the 179.5 MHz channel at time integrations of 13, 22, and 23. The significant contrast between the phase of the sky fringe, and how it's restricted to a narrow-band is an obvious indication of being RFI.}
\label{fig:AmpPhsExample}
\end{figure}

In this paper we expand upon the RFI identification approach using a DFCN developed in \texttt{Tensorflow} \citep{Tensorflow2016} with the use of both the amplitude and phase information from an interferometric visibility. This technique is prompted by examples such as what is shown in Figure \ref{fig:AmpPhsExample}, which demonstrates how the phase of time-ordered visibilities (waterfall visibilities) can provide supplemental information in identifying RFI beyond that of an amplitude-only approach. Note that in this paper, all time-ordered visibility plots of real data are in the yellow-purple palette (e.g. Figure \ref{fig:AmpPhsExample}) whereas all simulated data is in the blue-white palette (e.g. Figure \ref{fig:PredictionComparisonSim}). To understand the improvements afforded by our joint amplitude-phase network we compare it to both an amplitude only network and the Watershed RFI algorithm (See Appendix~\ref{Appendix:A}) which is the current RFI-flagging algorithm of choice for the HERA data processing pipeline.

The paper is outlined as follows. Section~\ref{sec:method} introduces the architecture of our network, discusses how it compares to previous work, and describes the training dataset. We then demonstrate the effectiveness by evaluating both DFCNs on simulated and real HERA observations in Section~\ref{sec:eval}. Finally in Section~\ref{sec:conclusions} we conclude with discussion of further applications and future work.

\section{Method}
\label{sec:method}
\subsection{DFCN Architecture}

The standard 2d convolutional neural network (CNN) \citep{LeCun1998_1} is structurally similar to that of a typical Artificial Neural Network (ANN) \citep{LeCun1998_2}, but it differs to an ANN's dense layers of `neurons' by its successive convolutions of an input image, which preserves spatial dependence. Each convolutional layer contains a set of learn-able filters which represent a response for particular shapes at different scales (e.g. the edges of an object in an image). The convolved output for every layer is then typically downsampled using a process known as max pooling that strides a window across the image keeping the highest pixel value within the window. Max pooling provides both a computational improvement due to a decreased image size, and an added level of abstraction relative to the initial image. After the convolution and max pooling layers the image typically is then passed through a non-linear activation function (e.g. sigmoid function) which produces a spatial activation map describing the convolutional layer's response to every pixel contained within the image. The eventual output of these successive convolution, max pooling, and activation layers is then used to predict (or regress) based on the classification of the image. The error between the predicted class and the true class is then computed through a loss function such as the cross-entropy loss (or mean squared error for regression) and the error is back-propagated through the network updating the learn-able parameters.

The style of network we describe in this paper deviates from a traditional CNN by requiring a fully connected convolutional layer of neurons after the convolutional downsampling and an upsampling stage to semantically predict classes on a per-pixel basis. For a deeper understanding of this kind of network architecture, see \cite{Krizhevsky2017}. We begin with a Deep Fully Convolutional Network architecture similar to the U-Net RFI \citep{Ronneberger2015,Akeret2017} implementation. However, instead of using a uniform number of feature layers for each convolutional layer, we use an image pyramid \citep{Lin2016} style approach with an increasing number of features as the network approaches the fully connected convolutional layers and invert to a decreasing number of feature layers towards the output prediction layer. This approach should offer us an increase in performance as the input image for each successive convolution shrinks. Each stacked layer in the max pooling stages has the dimensions of 
$(\frac{H}{2^{L}} \times \frac{W}{2^{L}} \times 2^{L}F)$ where $F$ is the number of feature layers, $H$ and $W$ are the layer height and width in pixels, and $L$ is the layer of interest.

\begin{figure*}
\includegraphics[scale=.6]{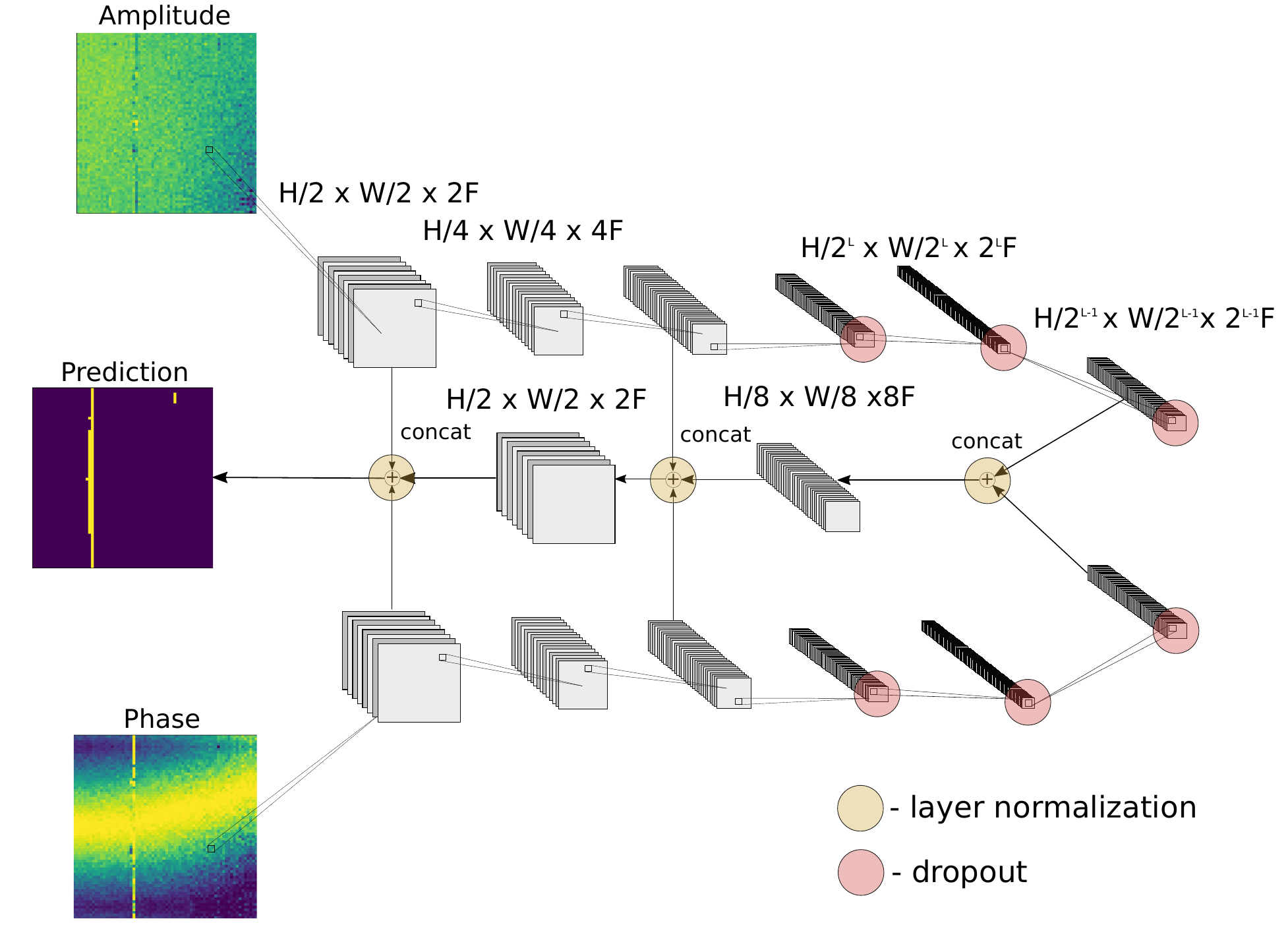}
\caption{The general architecture for the amplitude-phase DFCN demonstrating the sliced in frequency, padded in both time and frequency, and finally normalized amplitude \& phase input layers. H and W correspond to the input visibility dimensions in time and frequency, while F is the number of filter layers with L corresponding to the total number of layers between input and the fully convolutional layer. For reasons explained in Section \ref{sec:dataprep}, we use layer normalization at each skip connection and concatenation due to the difference in distributions of the amplitude and phase downsampling pathways. Every convolutional layer in the downsampling pathway is a 3$\times$ stacked set of convolutional layers with $3\times 3$ kernels leading into an output convolutional layer with a $1\times 1$ kernel, similar to the `Network in Network' architecture of \protect \cite{Lin2013}.}
\label{fig:architecture}
\end{figure*}

To adapt the network to use the visibility phase component, we mirror the amplitude only network as shown in Figure~\ref{fig:architecture}. We then combine successive amplitude \& phase convolution layers at each transpose convolution layer with the technique known as `skip connections' introduced in \cite{Long2014} and \cite{He2016}. This is implemented by taking the output of a downsampled convolutional layer and concatenating it with an upsampled transpose convolutional layer of equal time, frequency, and feature dimensions. By using these skips in the convolutional pathway, the network is provided with an initial ``template'' from which to make small deviations. This fixes an issue within deep networks where fits to higher-order nonlinearities become dominant in a layer, leading to training and overfitting issues. Empirically, we find that using skip connections in conjunction with phase information allows for training a deeper network that converges in fewer iterations than the simple amplitude-only network.

\begin{table}
\begin{tabular}{lcccc}
\hline \hline
\textbf{layer type} & \textbf{\begin{tabular}[c]{@{}c@{}}kernel size\end{tabular}} & \textbf{stride} & \textbf{filters} & \textbf{depth} \\ \hline
convolution & 3x3 & 1 & 16 & 2 \\ \hline
convolution & 1x1 & 1 & 16 & 1 \\ \hline
maxpool & 2x2 & 2 &  & 1 \\ \hline
batch norm. &  &  &  &  \\ \hline
convolution & 3x3 & 1 & 32 & 2 \\ \hline
convolution & 1x1 & 1 & 32 & 1 \\ \hline
maxpool & 2x2 & 2 &  & 1 \\ \hline
\rowcolor[HTML]{FFCE93} 
batch norm. &  &  &  &  \\ \hline
convolution & 3x3 & 1 & 64 & 2 \\ \hline
convolution & 1x1 & 1 &  64 & 1 \\ \hline
maxpool & 2x2 & 2 &  & 1 \\ \hline
\rowcolor[HTML]{FFCCC9} 
batch norm. &  &  &  &  \\ \hline
convolution & 3x3 & 1 & 128 & 2 \\ \hline
convolution & 1x1 & 1 & 128 & 1 \\ \hline
maxpool & 2x2 & 2 &  & 1 \\ \hline
\rowcolor[HTML]{ECF4FF} 
batch norm. &  &  &  &  \\ \hline
convolution & 3x3 & 2 & 256 & 2 \\ \hline
convolution & 1x1 & 1 & 256 & 1 \\ \hline
maxpool & 2x2 & 2 &  & 1 \\ \hline
batch norm. &  &  &  &  \\ \hline
\rowcolor[HTML]{ECF4FF} 
transpose conv. & 3x3 & 2 & 128 & 1 \\ \hline
\rowcolor[HTML]{FFCCC9} 
transpose conv. & 3x3 & 2 & 64 & 1 \\ \hline
\rowcolor[HTML]{FFCE93} 
transpose conv. & 3x3 & 2 & 32 & 1 \\ \hline
transpose conv. & 5x5 & 4 & 2 & 1 \\ \hline \hline
\end{tabular}
\caption{Architecture overview of the DFCNs demonstrated in this analysis. The colored rows correspond to the concatenations on the outputs between those respective layers, where prior to the concatenation each layer undergoes a batch normalization. The depth of a layer here means that there are multiples of the layer stacked all having the same properties. The amplitude-phase DFCN has two input pathways mirrored up until the first transpose convolution layer.}
\label{table:arch}
\end{table}


For each of the skip layer concatenations between the amplitude and phase pathways, we subtract the mean and normalize over both time and frequency, which assists in standardization as amplitude and phase features can be quite dissimilar. The amplitude only DFCN we use has $\sim$ 6$\times 10^{5}$ trainable parameters, while the addition of the phase downsampling layers for the amplitude \& phase DFCN pushes the number of trainable parameters to $\sim$9$\times 10^{5}$. The specific per layer attributes employed in our networks can be seen in Table \ref{table:arch}, where it should be noted that per layer dimension sizes are not specified because this style of network is agnostic to the input height and width.

To optimize the network hyperparameters, a coarse grid search was performed over dropout rate, learning rate, and batch size; the optimal results from this search are found in Table \ref{table:ArchitectureParams}. The depth of our convolutional layers are chosen to maximize learning and minimize prediction times, while trying to retain abstractions of the input visibilities that can properly describe our RFI. These dimensions are thus determined by initially training at an arbitrarily high number of feature layers and scaling back to the minimum number of layers we need to retain for convergence of the training loss.

\subsection{Data Preparation}
\label{sec:dataprep}
The analysis in this paper is performed entirely on HERA data (both simulated and real) and therefore should be noted that any data preparation techniques outlined here may be unique to HERA. This does not imply that they are unsuitable for other radio interferometers but additional precautions may need to be taken into consideration.
To prepare the amplitude-phase input space to be as robust to as many visibility scenarios as possible, we must adopt several standardizations. The amplitude of the visibility can vary drastically by local sidereal time (LST), day, and baseline type while having significant differences in dynamic range. In contrast, the phase of a visibility is intrinsically more standardized: it is constrained between $-\upi \leq \phi \leq \upi$ and should have a mean that is approximately $\mu_{\phi} = $ 0, so we should only expect substantial deviations across baseline type, which are due to changing fringe rates. Therefore to lessen the dynamic range issues in amplitude, we standardize our waterfall visibilities $V(t,\nu)$, according to $\hat{V}(t,\nu) = (\mathrm{ln}|V| -\mu_{\mathrm{ln} |V|})/\sigma_{\mathrm{ln} |V|}$, by subtracting the mean, $\mu_{\mathrm{\mathrm{ln} |V|}}$, and dividing by the standard deviation, $\sigma_{\mathrm{\mathrm{ln} |V|}}$, across time and frequency of the logarithmic visibilities.

To further increase the robustness and generalizability of our network for different time and frequency sub-bands, we slice the HERA visibilities into 16 spectral windows of dimensions 64 frequency channels by 60 time integrations ($6.3 \ \mathrm{MHz} \times 6 00 \ \mathrm{sec}$). We then pad both time and frequency dimensions by reflecting about the boundaries, extending the dataset in both directions. This allows for making predictions for the edge pixels, which otherwise would be ignored due to the size of our convolution layer kernel size of $3 \times 3$ ($98.44\ \mathrm{kHz} \times 30\ \mathrm{s}$). Furthermore, we want to use square input channels to maintain a 1:1 aspect of time to frequency pixels. These considerations inform the decision to use a 68$\times$68 input image.

Combined with our training batch size, $N$, of 256, for our amplitude-phase DFCN, this gives us an input training space of size $N\times H\times W\times C = (256\times 68\times 68\times 2)$, where C is the number of input channels (e.g. $C_{0}$, $C_{1}$ = $\hat{V}(t,\nu)$, $\phi(t,\nu)$).

\subsection{Training Dataset}
\label{sec:trainingdataset}

\begin{table}
\caption{Parameters and network architecture features that were determined by grid-search cross validation. The dropout rate is uniform across all nodes as highlighted in Figure \ref{fig:architecture}.}
\label{table:ArchitectureParams}
\begin{tabular}{lc}
\hline
Parameter & Values \\
\hline
Batch size & 256 \\
Optimizer & ADAM$^1$ \\
Learning rate & 0.003\\
Activation function & Leaky Rectified Linear Unit$^2$\\
Dropout rate & 0.7\\
Loss Function & Cross Entropy\\
\hline
\multicolumn{2}{l}{$^1$\citet{Kingma2014}} \\
\multicolumn{2}{l}{$^2$\citet{Maas13}}
\end{tabular}
\end{table}

The training dataset was composed of simulated HERA visibilities using the simulator, hera\_sim.
\footnote{\url{https://github.com/HERA-Team/hera\_sim}}.

This simulator creates visibilities according to a `pseudo-sky', which means that modeled point sources have no relationship, in either time or frequency, to any real extragalactic source on the sky (e.g. Fornax A). Extragalactic point sources are modeled using the discrete form of the visibility equation
\begin{equation}
 \tilde{V}(t,\nu) = \sum_{n}\Big[\tilde{A}(\tau,\hat{s})*\tilde{S}_{n}(\tau)*\delta(\tau_{n} - \tau) \Big] 
\end{equation} 
\citep{Parsons2012} which depends upon the source delay position on the sky, $\tau$, the source spectrum, $S_{n}(\nu)$, and the delay-dependent interferometer gains, $\tilde{A}(\tau,\hat{s})$, where a tilde represents the Fourier transform converting between frequency $\nu$ and delay $\tau$ and $*$ represents a convolution. Point source flux densities are drawn from a power-law distribution of the form $\mathrm{Pr}(S > S_{\small 0.3 \ \mathrm{Jy}}) = \Big(\frac{S}{S_{\small 0.3 \ \mathrm{Jy}}}\Big)^{-1.5}$ with a lower bound of 0.3 Jy. The spectral indices for these sources are then assigned uniformly at random between $-1 \leq \alpha_{r} \leq -\frac{1}{2}$ as per \citet{2017HurleyWalker}, where $S_{\nu} \propto \Big(\frac{\nu}{\nu_{center}}\Big)^{\alpha_{r}}$. The source delays (sky positions) are also chosen according to a uniform random distribution. Each simulated waterfall visibility contains between $10^{3} \leq N_{srcs} \leq 10^{4}$ sources with the aforementioned characteristics. We simulate diffuse galactic emissions with the \citet{Oliveira2008} Global Sky Model (GSM) and an analytic form of the HERA primary beam \citep{ParsonsMemo2015}. GSM diffuse emissions are not precisely modeled but created to give a sky-like realization by sampling across LST and frequency, and applying a filter in time that has a fringe-rate corresponding to the baseline type being simulated. The visibility baseline types are uniformly sampled across LST, where baseline length, $|\vec{b}|$, is chosen according to a half-normal distribution with $\mu_{{|\vec{b}|}} = 7.5 \ \lambda$  and $ \sigma_{|\vec{b}|} = 150 \ \lambda$. This is done to closely resemble the distribution of baseline lengths seen in HERA which is weighted towards short baselines. The learned model can be further tuned as longer baseline types are introduced.

We model RFI with four distinct classes: narrowband persistent (e.g. ORBCOMM), narrowband burst (e.g. ground/air communications), broadband burst (e.g. lightning), and random single time-frequency `blips'. Narrowband persistent constitutes the majority of RFI and are most often the brightest sources in HERA observations; these are empirically modeled. Narrowband bursts have no preference in duration or frequency but typically persist > 30 s and are simulated with a Gaussian profile in time to mimic the roll on/off seen in HERA observations. Broadband bursts are rare events that exist across the entire HERA band at specific time integrations. These events are introduced in only 3\% of the training data. We randomly inject `blips' that are RFI with a duration of  $\Delta t \leq$ 10 s and frequency width,  $\Delta \nu \leq$ 100 kHz, which when taking into account HERA's time and frequency resolution places this class of RFI into single visibility pixels.

To create the most comprehensive HERA visibility simulations to mimic real observations we include simplistic models of several important effects seen in the HERA signal chain. These effects include: 

\begin{description}
\item \textbf{Cross-talk} - An effect due to over-the-air coupling between nearby HERA receivers and dipole-arm coupling. This spurious correlation is mocked by convolving the simulated visibility with white noise.\\
\item \textbf{HERA bandpass} - Empirically derived from HERA bandpass measurements and fit to a $7^{th}$ order polynomial \citep{Parsons2017}. \\
\item \textbf{Gain fluctuations} - Fluctuations are applied to the analytic HERA bandpass by introducing individual phase delays with a uniform spread between $-20 \leq \delta \tau \leq 20$ ns. \\
\end{description}

\begin{figure}
\includegraphics[scale=2.5,width=\columnwidth]{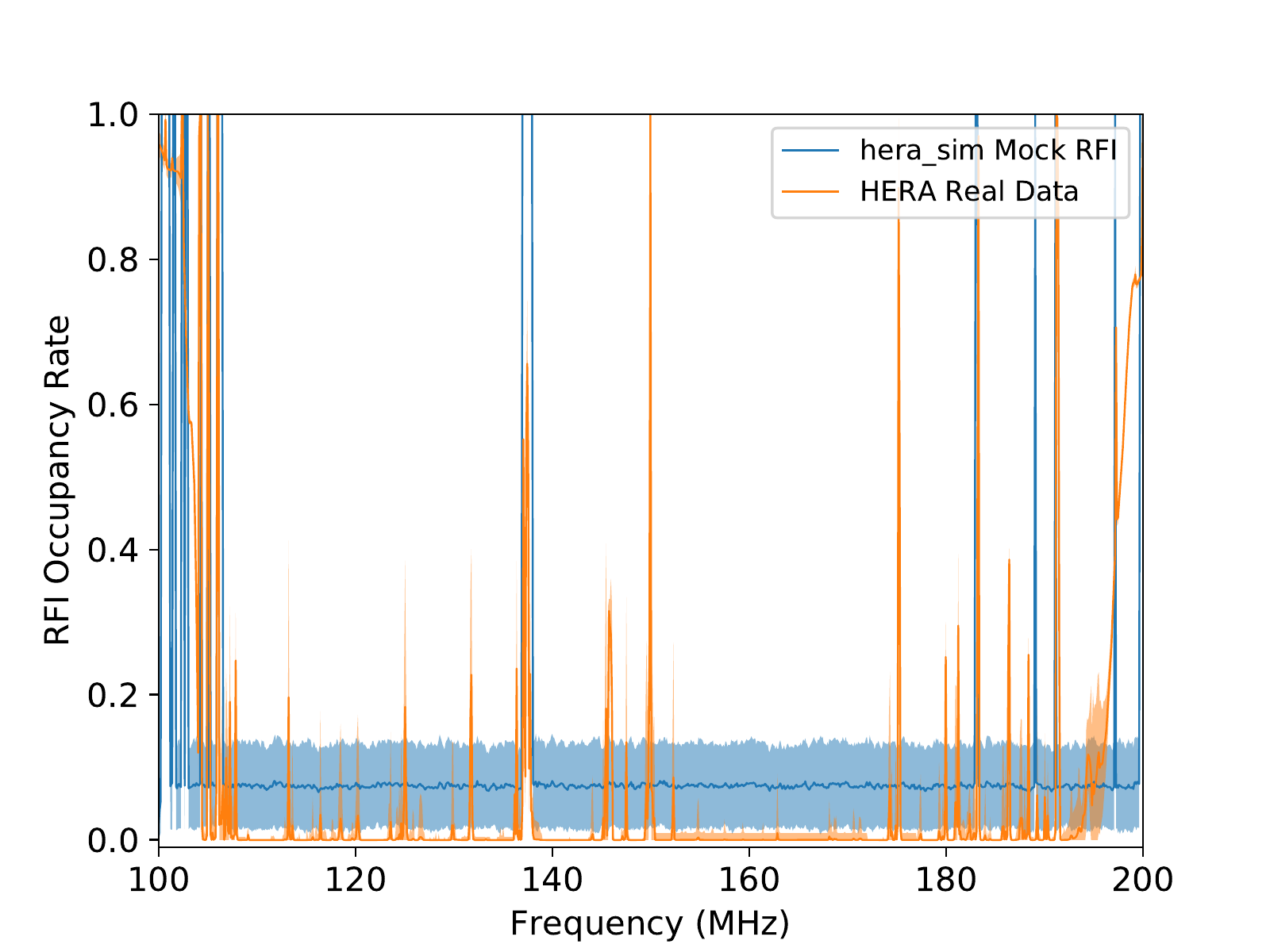}
\caption{The hera\_sim mock RFI (blue) occupancy rates across the band with its variance (blue region), as compared to RFI flagged in our real HERA data evaluation dataset (orange) with its own variance (orange region). The simulated RFI is overemphasized (> 10 \%) in the training dataset. This is done in an attempt to balance the training due to RFI being a significantly sparse class which without would lead to more significance placed on Non-RFI when computing the loss.}
\label{fig:SimOccupancy}
\end{figure}

We simulate a training dataset of 1000 HERA observations of 10 minutes (60 time integrations) over the frequency range of 100-200 MHz (1024 frequency channels). The mean RFI occupancy rate for these simulated observations was $\sim$ 10\%. This value differs from the $\sim$3\% which is the typically observed RFI environment in the Karoo Desert, South Africa seen in past HERA observations \citep{Kohn2016} which used a simple statistical thresholding RFI algorithm. The comparison between our simulated and more recent real HERA dataset RFI occupancy rates across the band can be seen in Figure \ref{fig:SimOccupancy}. We further expand this training dataset by performing data augmentation techniques on the reduced spectral windows. These techniques include mirroring over time and frequency, Gaussian random noise injection  (correlated between amplitude and phase) with an amplitude that is at most 10\% of $|V|$ the visibility amplitude and by translating a spectral window across the band creating unique window samples at varying frequency intervals. Using a translation in frequency has the intent of reducing over-fitting to steady state narrow-band RFI (e.g. ORBCOMM) because of repetitive sub-band samples entering the training dataset. 

 After increasing our simulated dataset volume through augmentation it is sliced into 16 spectral windows and padded according to Section \ref{sec:dataprep} which results in 44800 unique spectral window visibilities each of size 68 time samples $\times$ 68 frequency channels. We separate this simulated dataset by an 80-20 split, where 80\% of the simulated dataset is used for training and 20\% is used for validation.

\section{Evaluation}
\label{sec:eval}
For the evaluation of our networks we used several datasets unseen in training. The real observed dataset used for evaluation consisted of HERA observing data from the 2017 - 2018 season, more specifically between the Julian Dates of 2458098 - 2458116, which we will just refer to as our real HERA dataset . The real HERA data were composed of raw uncalibrated visibilities that have been  visually inspected and manually flagged by hand with high- and low-frequency band edges removed.  Hand flagging was accomplished by looking in both amplitude and phase for sharp discontinuities and structure that exhibited an increase in power when compared to a fringing sky signal. The band edge removal is a precaution due to the large dynamic range roll-off, which makes discriminating between RFI and sky observations nearly impossible for humans and algorithms alike. This reduces our actual data evaluation passband to 896 frequency channels which covers $106 \leq \nu \leq 194$ MHz.

A simple approach to evaluation would rely on using the accuracy of predictions meaning we only look at the number of correctly predicted RFI pixels relative to all RFI, although this metric hides important details to the performance of our networks. This is an important consideration in this instance because our HERA data observations contain on average 3\% of data corrupted by RFI, which means a blanket classification of ``No-RFI'' would yield an accuracy rate of 97\%. To account for this class imbalance we evaluate the effectiveness of our networks by using several metrics commonly employed for classification. We use the standard metrics of Recall and Precision, which are defined as
\begin{equation}
\mathrm{Recall} = \frac{\mathrm{TP}}{\mathrm{TP + FN}} = \frac{\mathrm{RFI}_\mathrm{Correct}}{\mathrm{RFI}_\mathrm{Correct} + \mathrm{RFI}_\mathrm{Incorrect}}
\end{equation}
\begin{equation}
\mathrm{Precision} = \frac{\mathrm{TP}}{\mathrm{TP + FP}} = \frac{\mathrm{RFI}_\mathrm{Correct}}{\mathrm{RFI}_\mathrm{Correct} + \mathrm{NoRFI}_\mathrm{Incorrect}}
\end{equation}
where we consider True Positives (TP) to be correctly predicted RFI pixels, False Negatives (FN) are RFI pixels identified as No-RFI, and False Positives (FP) are No-RFI pixels incorrectly identified as RFI. We can therefore understand Recall as the fraction of all RFI events identified by the flagging algorithm, and Precision the fraction of identified RFI that is actually RFI.

\begin{table*}
\caption{RFI recovery metrics based on individual type of simulated RFI. We look at the Recall, Precision, and $F_{2}$ score for each of the three algorithms as simulated with hera\_sim. The Recall and Precision rates are the average over 1000 simulated waterfall visibilities with the same simulation parameters for foregrounds and signal chain outlined in Section \ref{sec:trainingdataset}. Values in bold indicate the best achieved rate within each RFI type across algorithms.}
\label{table:RecoveredTypes}
\begin{tabular}{lccccc}
\hline
& & & RFI Classes  & & \\
\cline{2-6}
  & No RFI & Narrowband & Narrowband Burst & Broadband Burst & `Blips'\\
 \ \ \ \ Algorithm & Accuracy (\%) & Recall - Precision - $F_{2}$  & ``         '' & ``         ''  & ``         '' \\
\hline
Amp DFCN & 94 & 0.98 - 0.82 - 0.94 & \textbf{0.77} - 0.65 - \textbf{0.74} & 0.16 - 0.67 - 0.19 & 0.35 - 0.01 - 0.07\\
Amp-Phs DFCN & \textbf{98} & \textbf{0.99} - \textbf{0.83} - \textbf{0.95} & \textbf{0.77} - 0.67 - \textbf{0.74} & 0.18 - 0.68 - 0.21 & 0.35 - 0.02 - 0.08 \\
Watershed RFI & \textbf{98} & 0.49 - \textbf{0.95} - 0.54 & 0.32 - \textbf{0.97} - 0.37 & \textbf{0.99} - \textbf{0.74} - \textbf{0.98} & \textbf{0.71} - \textbf{0.73} - \textbf{0.71} \\
\hline
\end{tabular}
\end{table*}

We also need a metric that is sensitive to a dataset with a sparse class, as in our case where RFI represents $< 3\%$ of our observations, and one that can condense our overall performance into a single metric. We therefore can use the binary classifier performance metric, the F-score which has the general form of
\begin{equation}
F_{\beta} = (1 + \beta^2) \ \frac{\mathrm{Recall}\cdot\mathrm{Precision}}{\beta^2 \cdot \mathrm{Precision} + \mathrm{Recall}}
\end{equation}
where we set $\beta = 2$ to preferentially weight Recall\footnote{An $F_{\beta}$ score with $\beta < 1$ describes a preference for weighting Precision over Recall.} The $F_{2}$ score therefore provides us with a metric that has an aggressive stance towards RFI while still being somewhat sensitive to false positive flagging.

Due to the nature of measuring the 21cm EoR signal with HERA where we have collected sufficient data to not be noise limited, we can sacrifice good quality observations for the sake of reducing as much RFI contamination as possible; thus allowing for a higher rate of FPs. This leads us to maximizing Recall while allowing Precision to suffer which is preferential when averaging over many nights of observations and we want to minimize contamination.

We compare three distinct algorithms: the amplitude-only DFCN, amplitude-phase DFCN, and the Watershed RFI algorithm. For a fair comparison we evaluate the DFCNs after both have converged independently of the number of training epochs, ensuring that they have learned the training dataset to their maximum capability. The networks are then checked for over-fitting by comparing that the evaluation loss is similar to that of each networks training loss when applied to the unseen 30\% of simulated visibilities set aside for evaluation.

\begin{figure*}
\includegraphics[scale=.6,width=\linewidth]{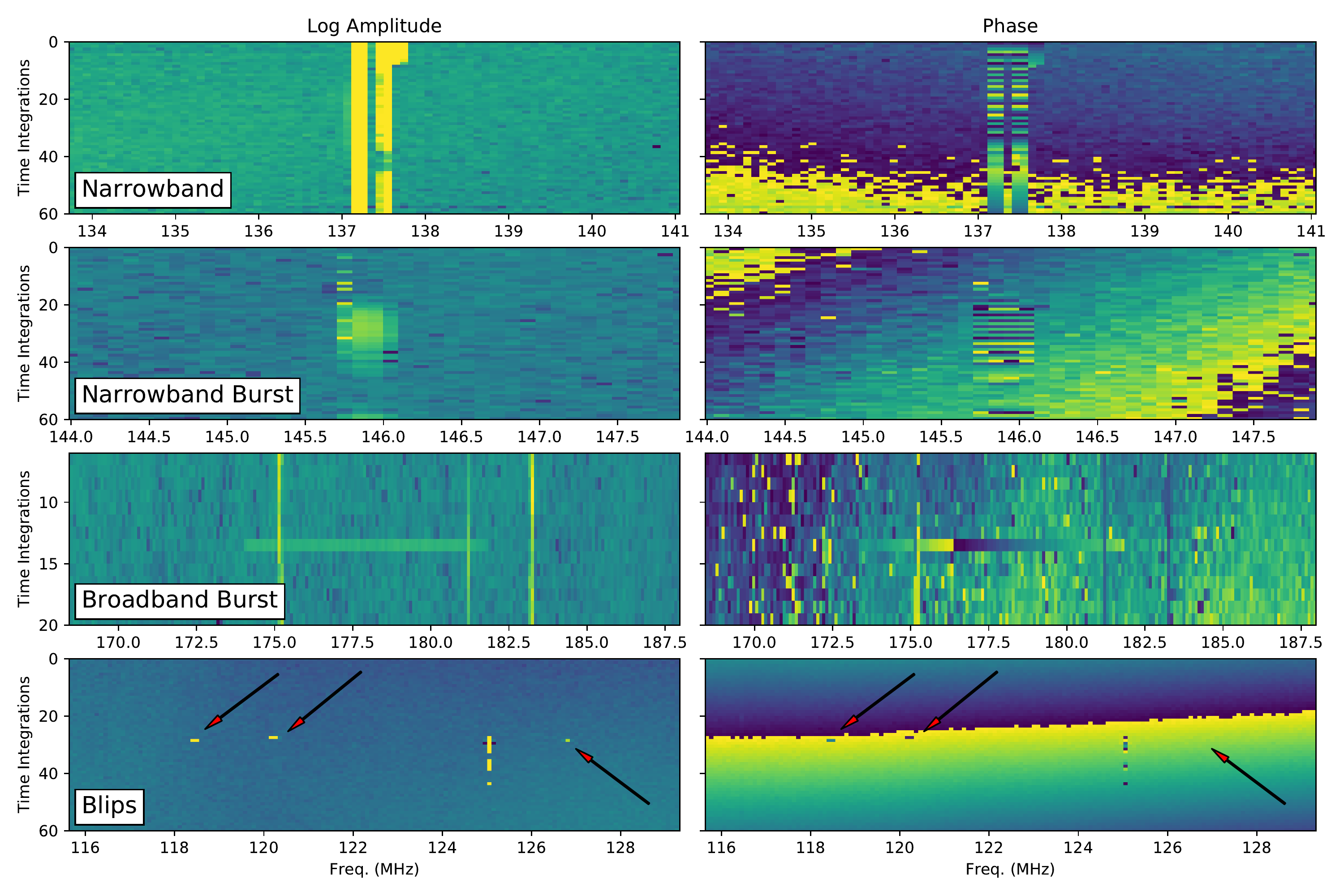}
\caption{Examples of the four RFI classes from HERA data as they appear in amplitude and phase that we model in our simulations. Note the different time and frequency scales on each plot. The narrowband example (row 1) centered at a frequency of $\sim 137.2$ MHz is the ORBCOMM satellite system which is occasionally intermittent. Narrowband burst (row 2) is typically limited across only a few frequency channels ($\leq 500$ kHz) and has no consistent operating pattern over time. Broadband burst events (row 3) are short time duration ($\leq 40$ s) and can exist across the entire band (e.g. lightning) or in a sub-band as seen here flanked by the South African Broadcasting Corporation's channel 4 video (175.15 MHz) and audio (181.15 MHz) broadcasts \citep{Kohn2016}. The `blips' (row 4) demonstrate the one off nature of this sparse class as compared to the intermittent transmitter at frequency 125 MHz.}
\label{fig:RFIClasses}
\end{figure*}

The results of each algorithm along with their performance metrics are reported in Table~\ref{table:RecoveredTypes} as applied to simulated datasets. These metrics are performed on data that are unique from the previous training/evaluation datasets and include a control dataset which has no RFI present and four others that contain a single distinct class of RFI.  An example of what each RFI class is modeled after is shown in Figure \ref{fig:RFIClasses}. In doing this, we can gauge how sensitive each algorithm is to a certain class of RFI. The DFCN networks both perform well on the narrowband time persistent and burst RFI which is unsurprising as these simulations closely resemble the evaluation dataset and only differ in occurrence of events. However, both networks are inadequate for identifying Broadband Bursts and `blips'. This is understandable from a training perspective as both of these classes of RFI are going to be the last to be modeled in our networks as they account for only a minor fraction of all simulated RFI and lead to little overall optimization of the loss. This could potentially be remedied by placing more emphasis on these two classes of RFI in the training dataset or a much more in depth hyperparameter optimization of per-layer kernel sizes.

\begin{figure}
\includegraphics[scale=1.5,width=\columnwidth]{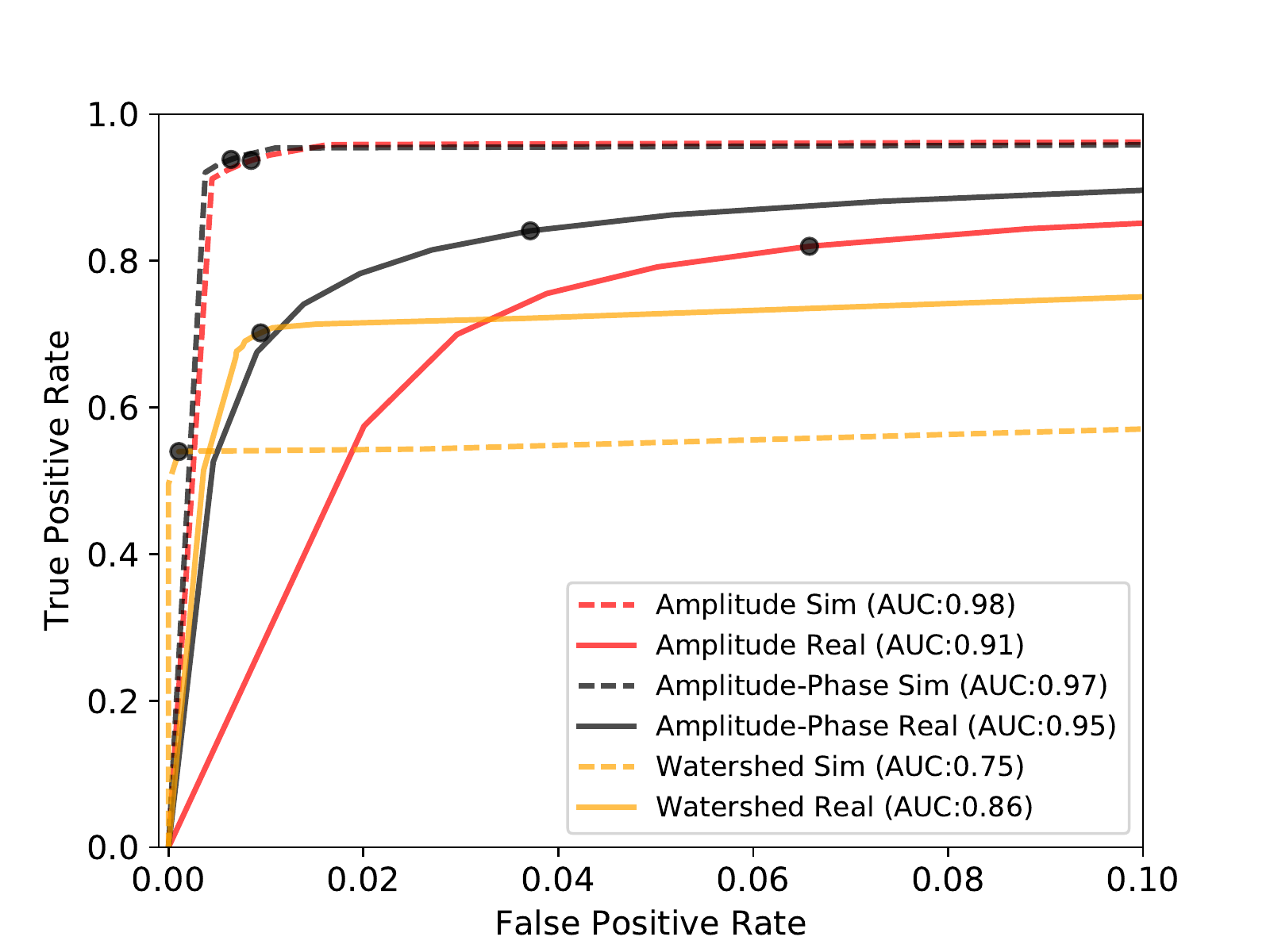}
\caption{ROC curve comparing all three RFI flagging algorithms, Amplitude DFCN (Red), Amplitude-Phase DFCN (Black), and Watershed (Orange). The ROC curves were derived from each algorithm predicting on real HERA data visibilities (solid) and simulated HERA visibilities (broken). Black circles represent the optimal $F_{2}$ score. The Area Under the Curve (AUC) metric condenses the overall performance of our algorithms and tells us that the Amplitude-Phase network exhibits the best response on our real data with an AUC of 0.95. The TPR and FPRs for the real data (solid) are based on manually flagging RFI to the best of our ability to discern RFI from signals on the sky and therefore should not be taken as a ground truth.} 
\label{fig:ROC}
\end{figure}

Before we approach the evaluation on our HERA data data we further optimize how our networks handle the shift in domain from simulation to observed. This is done by looking at the Receiver Operating Characteristics (ROC) curves in Figure \ref{fig:ROC} of both the networks and the Watershed algorithm. The ROC curve gives us an idea of the performance of our network by looking at how the True and False Positive rates respond to different thresholding values of the networks' softmax output layer.

\begin{figure*}
\includegraphics[scale=.6,width=\linewidth,trim={6cm 0 0 0},clip]{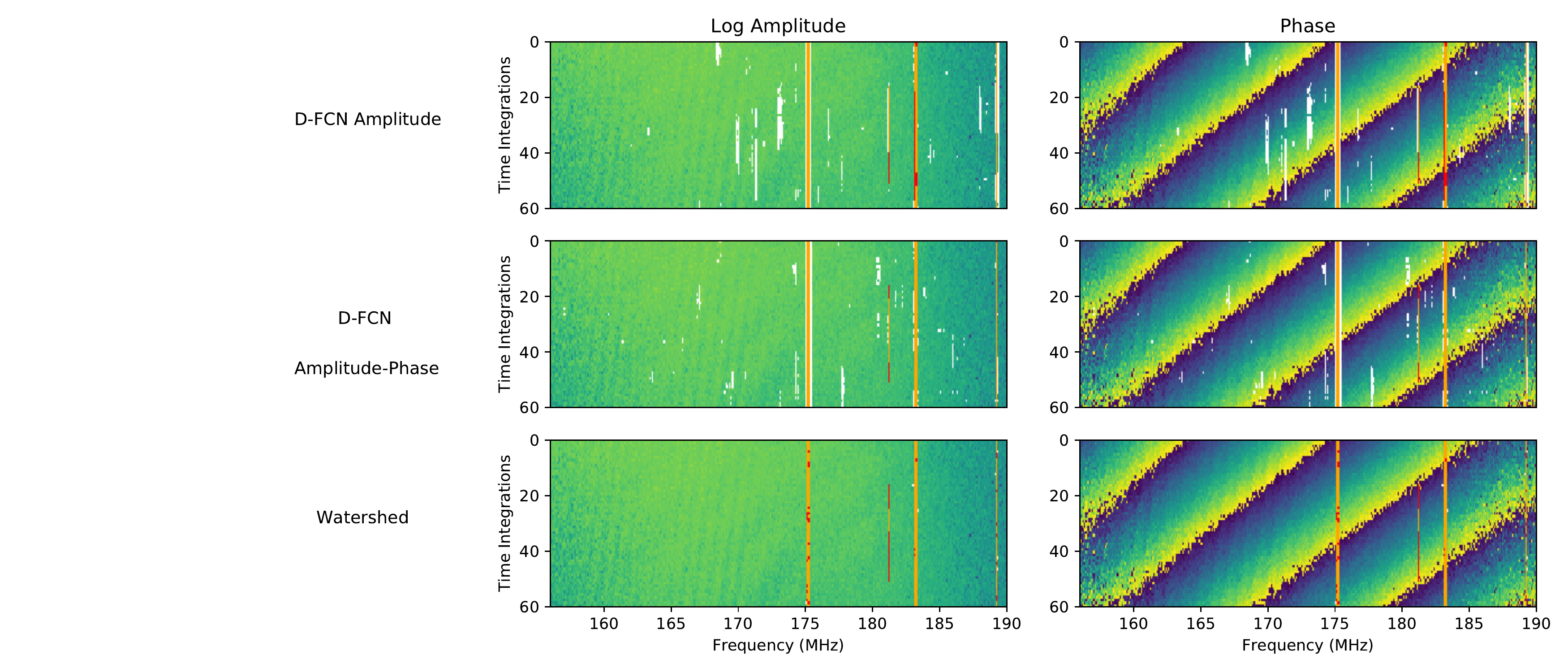}
\caption{A comparison between the three flagging algorithms described in this paper as applied to a sub-band (157 - 193 MHz) from the real HERA dataset, which has been flagged manually and has no known ground truth. Orange indicates true positives, white is false positives, and red represents false negatives. In this example the amplitude-phase fed DFCN ultimately has the best true positive outcome but, as seen in Table \ref{table:RealData}, both the DFCN algorithms take a more aggressive stance towards RFI resulting in higher rates of false positives when compared to the Watershed algorithm.}
\label{fig:PredictionComparison}
\end{figure*}

\begin{figure*}
\includegraphics[scale=.6,width=\linewidth,trim={6cm 0 0 0},clip]{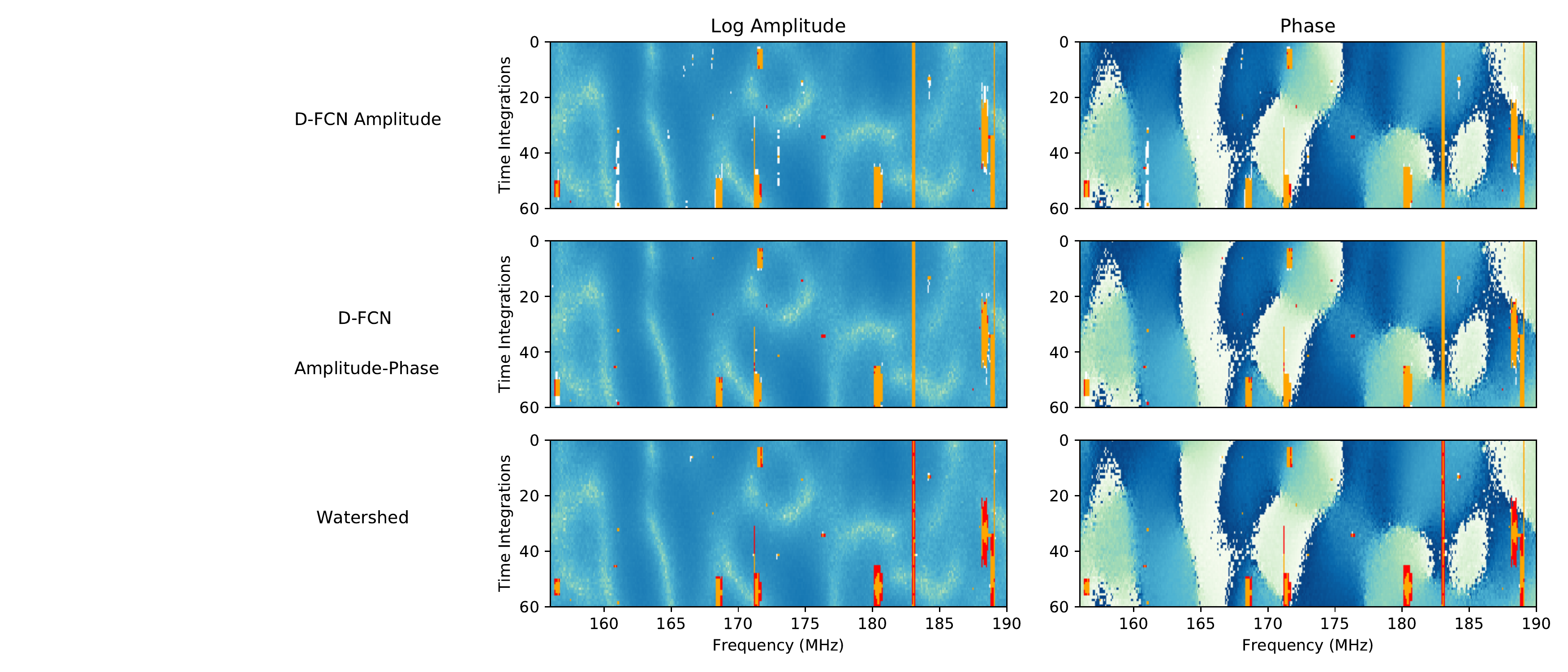}
\caption{A similar comparison as in Figure \ref{fig:PredictionComparison} demonstrating how each RFI flagging algorithm performs on simulated HERA data from hera\_sim. Orange indicates True Positive, white is False Positive, and red is False Negative. The simulated waterfall visibility is of a 25$\lambda$ baseline dominated by strong diffuse emissions from the \citet{Oliveira2008} GSM. The Watershed algorithm's inability to discriminate between RFI and sky, as indicated by its higher false negative rate, in this instance hints that there is required fine-tuning of its kernel size and initial threshold hyperparameters, due to the spectral structure in our simulations.}

\label{fig:PredictionComparisonSim}
\end{figure*}

We determine the optimal thresholding value by using the maximum $F_{2}$ score across all thresholds  which is shown to find a reasonable balance by locating the `knee' of the ROC curve. We then compare all three algorithms as applied to a simulated hera\_sim and real HERA dataset in Table \ref{table:RealData}. Looking at the prediction rates, both DFCN networks display an immense improvement over the Watershed RFI algorithm, boasting rates of 32$\times$ and 22$\times$ better than the Watershed, for the amplitude and amplitude-phase networks respectively. The faster amplitude only prediction rate compared to the amplitude-phase is unsurprising, as the number of parameters involved in an amplitude-phase prediction is roughly 1.5$\times$ more and scales approximately proportional with the prediction rate.
An example of these results, which serves to give an appropriate idea of the average performance as applied to a real HERA waterfall visibility is shown in Figure \ref{fig:PredictionComparison}; how each compares on simulated data can be seen in Figure \ref{fig:PredictionComparisonSim}. Both DFCN networks have a tendency to over-predict RFI where there may not be any, however in the case of the narrowband RFI seen in frequency channels 175 MHz and 189 MHz, it may not be unreasonable to be more aggressive as leakage into adjacent channels can occur. This can be difficult to quantify of course as the ground truth of our real HERA data is unknown and RFI leakage can be easily masked by the sky.

\begin{table*}
\caption{RFI recovery metrics for hera\_sim simulated data containing signal chain effects with all classes of RFI and raw (uncalibrated) HERA observations from the 2017 - 2018 observing season. All results in the real HERA data column are based off of manually identified RFI  and therefore the ground truth is uncertain especially in the low SNR limit for RFI. Our real HERA data included observations from LSTs of $0 \leq t \leq \ 5 \ \mathrm{h}$ and across baseline lengths of $ 7 \leq |\vec{b}| \leq 100 \ \lambda$. Values in bold correspond to the best achieved result for that metric.}
\label{table:RealData}
\resizebox{\width}{!}{%
\begin{tabular}{lccc}
\hline
 & hera\_sim & HERA real & Prediction Rate \\
Algorithm & Recall - Precision - $F_{2}$ & `` '' & waterfall/h/GPU\\
\hline
Amp DFCN & \textbf{0.90} - 0.61 - 0.82 & 0.76 - 0.42 - 0.65 & 2.4$\times 10^{5}$\\
Amp-Phs DFCN & \textbf{0.90} - 0.82 - \textbf{0.88} & \textbf{0.81} - 0.58 - \textbf{0.75} & 1.6$\times 10^{5}$\\
Watershed RFI & 0.53 - \textbf{0.95} - 0.58 & 0.64 - \textbf{0.88} - 0.68  & 7.4$\times 10^{3}$\\
\hline
\end{tabular}
}
\end{table*}

\section{Conclusions}
\label{sec:conclusions}
Machine learning applications in the fields of astronomy and cosmology are rapidly developing, and in many cases are beginning to outmaneuver the classical algorithms by way of increased speed and more accurate predictions. In this paper we described an RFI identification approach to using a Deep Fully Convolutional Neural Network, which combined the amplitude and phase of an interferometer's measured visibility to predict which time-frequency pixels contained RFI. We compared this result to the Watershed RFI algorithm in the HERA data processing pipeline, and demonstrated that the DFCN approach was vastly more time efficient in its prediction with comparable to improved RFI identification rates. We also show that by including the phase component of the visibility we can mitigate the effects of domain shift between an entirely simulated HERA visibility training dataset and the observed validation dataset. This means that by improving our simulated model for HERA visibilities, coupled with an amplitude-phase DFCN we should be able to achieve an extremely effective first-round RFI flagger that reduces a common pipeline bottleneck. We do however recognize that the DFCN approach can have issues with identifying RFI bursts that occupy single time-frequency samples, what we called `blips', and broadband bursts. This is most likely due to an imbalanced representation in our training dataset, and the loss optimization not being rewarded enough to drive the DFCNs to learn a subclass that appears at a rate of < 0.1\%. This can be potentially overcome by fine-tuning the model by using transfer learning \citep{Yosinski2014}, and would involve a training dataset which consists almost entirely of these two subclasses, where the trained DFCN model shown here would serve as the starting point.

In near future build-outs of HERA there will need to be an extreme importance placed on reducing bottlenecks in the HERA data processing pipeline. The current Watershed RFI flagging algorithm does not scale particularly well, which puts this class of fully convolutional neural network as an ideal alternative. The eventual number of HERA dishes will total 350, which for a single 10 minute observation gives us 61,075 unique waterfall visibilities. In the amplitude-phase DFCN design outlined in this paper the RFI flagging throughput is 1.6$\times 10^{5}$ waterfalls/h/gpu \footnote{Performed on a single NVIDIA GeForce GTX TITAN} which compares to the Watershed RFI flagger at 7.4$\times 10^{3}$ on the same resources.

Future work related to the amplitude-phase DFCN could include a modification to a similarly styled comprehensive data quality classifier which should in-turn lead to improved results for sky based \citep{Barry2016} and redundant calibration \citep{Zheng2014}, both of which requires exceptionally conditioned data. A strict binary classifier could be achieved by developing a training dataset that doesn't use a mock sky, but an accurately modeled sky with a proper HERA beam model. Of course it would also be possible and might be better suited by developing an observation derived training dataset in this instance, as failure modes are generally easier to identify in visibilities as opposed to contamination by RFI.

It should also be possible to extend this work to arrays with better temporal resolution such as the MWA \citep{Tingay2015} in the search for transients like Fast Radio Bursts (FRBs, \citealt{Zhang2018}). The additional phase information could potentially reduce the low-end limit of fluence for identification due to a more significant contrast between RFI and sky fringes. 

The github repository for the RFI DFCN described in this paper can be found at \url{https://github.com/UPennEoR/ml\_rfi}.

\section*{Acknowledgements}
This material is based upon work supported by the National Science Foundation under Grant Nos. 1636646 and 1836019 and institutional support from the HERA collaboration partners.  This research is funded in part by the Gordon and Betty Moore Foundation. HERA is hosted by the South African Radio Astronomy Observatory, which is a facility of the National Research Foundation, an agency of the Department of Science and Technology.  This work was supported by the Extreme Science and Engineering Discovery Environment (XSEDE), which is supported by National Science Foundation grant number ACI-1548562 \citep{xsede2014}. Specifically, it made use of the Bridges system, which is supported by NSF award number ACI-1445606, at the Pittsburgh Supercomputing Center \citep{bridges2015}. We gratefully acknowledge the support of NVIDIA Corporation with the donation of the Titan X GPU used for this research. SAK is supported by a University of Pennsylvania SAS Dissertation Completion Fellowship. Parts of this research were supported by the Australian Research Council Centre of Excellence for All Sky Astrophysics in 3 Dimensions (ASTRO 3D), through project number CE170100013. GB acknowledges support from the Royal Society and the Newton Fund under grant NA150184. This work is based on research supported in part by the National Research Foundation of South Africa (grant No. 103424). GB acknowledges funding from the INAF PRIN-SKA 2017 project 1.05.01.88.04 (FORECaST). We acknowledge the support from the Ministero degli Affari Esteri della Cooperazione Internazionale - Direzione Generale per la Promozione del Sistema Paese Progetto di Grande Rilevanza ZA18GR02. This work is based on research supported by the National Research Foundation of South Africa (Grant Number 113121).





\bibliographystyle{mnras}
\bibliography{references}

\appendix

\section{Watershed RFI Algorithm}
\label{Appendix:A}
The current algorithm used in the HERA analysis pipeline is the Watershed RFI Algorithm, which performs some pre-processing of the raw data before identifying and removing suspected RFI instances. Before performing feature extraction, a median filter is applied to the data. In one dimension, a median filter is defined by the radius of the kernel $K$, which is applied as a sliding window across the entire length of the input data vector. Specifically, given an input vector $\va{x} = [x_0, x_1, \ldots, x_N]$, the median filtered output for a given entry $\tilde{x}_i$ can be expressed as:
\begin{equation}
\tilde{x}_i = \mathtt{median}(x_{i-K}, x_{i-K+1}, \ldots, x_{i-1}, x_i, x_{i+1}, \ldots, x_{i+K}),
\end{equation}
where \texttt{median()} is a function which returns the median of the list of data. By construction, the list will have an odd number of elements in it, and so the median is guaranteed to be an entry in the list.

In two dimensions, the median filter is defined analogously to the one dimensional case, except that there are two filter radii $(K_t, K_\nu)$ that define the median filter. Here we have used the subscripts $t$ and $\nu$ to represent the time and frequency axes found in a visibility waterfall. In general these need not be the same, but in practice as part of the HERA pipeline, both have the same value of $K_t = K_\nu = 8$. Empirically these values seem to fall into a ``sweet spot'' of parameter space, where the values were large enough that the overall algorithm catches the majority of RFI events (as verified by inspecting the visibilities by hand) while still remaining computationally tractable to run. Also, to ensure the output of the median filter has the same dimensionality as the input data, the arrays are padded with a reflection of the data that is $K_t$ or $K_\nu$ elements long, rather than with zero values, to avoid discontinuous jumps at the boundaries.

Physically, the median filter has the property of generating a proxy for the underlying noise of the raw visibility data because of its differencing of neighboring time-frequency pixels, and helps detrend the smooth foreground structure that is quite prominent and exhibits a strong frequency dependence. Once the two-dimensional median filter has been computed for every point in the visibility, the output is a ``noise'' visibility. The standard deviation of this ``noise'' is computed, which is then used to convert the noise to modified $z$-scores. (That is, the value of the noise is divided by the standard deviation, to quantify how strong of an outlier a particular data point is.) An initial round of seeds is generated by identifying all of the 6$\sigma$ outliers (the data points whose absolute valued $z$-scores is greater than six). Once the data has been pre-processed in this fashion, the watershed algorithm is used to identify all instances of RFI.

A watershed algorithm (or more correctly, a flood-fill algorithm, because the resulting image segments are not grouped or labeled) is then used to identify the remaining RFI instances in the waterfall\footnote{Please see \citet{Roerdink2000} for a more in-depth understanding of the Watershed algorithm}. Under the assumption that RFI events tend to have some coherency either in time (e.g., for narrow-band emission that is almost always on, such as ORBCOMM) or in frequency (e.g., for broad-band RFI events caused by lightning), the initial flags generated by finding 6$\sigma$ outliers are extended to neighboring pixels if the absolute value of their $z$-score is greater than 2. These regions are extended until no neighboring 2$\sigma$ values are encountered.

Algorithm~\ref{alg:watershed} shows the pseudocode of the XRFI flagging algorithm. The algorithm takes in a waterfall of visibility data $V_{ij}(t, \nu)$ and returns a set of flags $f_{ij}(t,\nu)$ of the same dimensionality. There are three main phases:
\begin{enumerate}
\item Pre-process the visibility data.
\item Generate initial series of flags.
\item Flood-fill around initial flags to generate full set of flags.
\end{enumerate}
As currently implemented, the watershed XRFI algorithm operates on the absolute value of the visibility data, though it could be extended to operate on the real and imaginary components as well. When running the watershed XRFI algorithm in production, the most computationally expensive part is the two-dimensional median filter which has a time complexity of $\mathcal{O}(K_t K_\nu)$. The overall complexity is roughly $\mathcal{O}(N_t N_\nu K_t K_\nu)$ for a waterfall visibility with dimensions $N_t \times N_\nu$. Thus, speeding up the median filter operation by decreasing the kernel size or leveraging GPU computing can provide a significant speedup.

\algblockx[Where]{Where}{EndWhere}{\textbf{where}~}{\textbf{end where}}
\algcblockx[ElseWhere]{Where}{ElseWhere}{EndWhere}{\textbf{else where}}{\textbf{end where}}
\begin{algorithm}
\caption{Watershed XRFI Algorithm}
\label{alg:watershed}
\begin{algorithmic}[1]
\Procedure{XRFI}{$V_{ij}(t,\nu)$}
    \State $\tilde{V}_{ij}(t,\nu) \gets \mathtt{medfilt2d}(V_{ij} (t,\nu), K_t, K_\nu)$
    \State $\sigma_{ij} \gets \qty(\sum_{t,\nu} \tilde{V}^2_{i,j}(t,\nu) - \sum_{t,\nu} \tilde{V}_{i,j}(t,\nu))^{1/2}$
    \State $z_{ij}(t,\nu) \gets \abs{\tilde{V}_{ij}(t,\nu) / \sigma_{ij}}$
    \Statex
    \Where $z_{ij}(t,\nu) > 6$\Comment{set initial flags}
        \State $f_{ij}(t,\nu) \gets \mathtt{True}$
    \ElseWhere
        \State $f_{ij}(t,\nu) \gets \mathtt{False}$
    \EndWhere
    \Statex
    \State $\mathtt{AddedFlags} \gets \mathtt{True}$
    \While{$\mathtt{AddedFlags}$}\Comment{flood fill to neighbors}
        \State $\mathtt{AddedFlags} \gets \mathtt{False}$
        \ForAll{$f_{ij}(t,\nu) \in t,\nu$}
            \If{$f_{ij}$ is \texttt{True}}\Comment{grow existing flags}
                \For{$t' \gets t \pm 1$}\Comment{check times}
                    \If{$z_{ij}(t', \nu) > 2$}
                        \State $f_{ij}(t', \nu) \gets \mathtt{True}$
                        \State $\mathtt{AddedFlags} \gets \mathtt{True}$
                    \EndIf
                \EndFor
                \For{$\nu' \gets \nu \pm 1$}\Comment{check frequencies}
                    \If{$z_{ij}(t, \nu') > 2$}
                        \State $f_{ij}(t,\nu') \gets \mathtt{True}$
                        \State $\mathtt{AddedFlags} \gets \mathtt{True}$
                    \EndIf
                \EndFor
            \EndIf
        \EndFor
    \EndWhile
    \Statex
    \State \textbf{return} $f_{ij}(t,\nu)$
\EndProcedure
\end{algorithmic}
\end{algorithm}

\section{Author Affiliations}
\label{appendix:affil}
$^{1}$Department of Physics, Brown University, Providence, Rhode Island, USA\\
$^{2}$Department of Physics and Astronomy, University of Pennsylvania, Philadelphia, Pennsylvania, USA\\
$^{3}$Department of Astronomy, University of California, Berkeley, CA\\
$^{4}$Cavendish Astrophysics, University of Cambridge, Cambridge, UK\\
$^{5}$SKA SA, 3rd Floor, The Park, Park Road, Pinelands, 7405, South Africa\\
$^{6}$School of Earth and Space Exploration, Arizona State University, Tempe, AZ\\
$^{7}$Department of Physics and Electronics, Rhodes University, PO Box 94, Grahamstown, 6140, South Africa\\
$^{8}$INAF-Istituto di Radioastronomia, via Gobetti 101, 40129 Bologna, Italy\\
$^{9}$National Radio Astronomy Observatory, Charlottesville, VA\\
$^{10}$National Radio Astronomy Observatory, Socorro, NM\\
$^{11}$Department of Physics, Massachusetts Institute of Technology, Cambridge, MA\\
$^{12}$Department of Physics and Astronomy, University of California, Los Angeles, CA\\
$^{13}$School of Physics, University of Melbourne, Parkville, VIC 3010, Australia\\
$^{14}$Department of Physics, University of Washington, Seattle, WA\\
$^{15}$eScience Institute, University of Washington, Seattle, WA\\
$^{16}$Department of Physics and McGill Space Institute, McGill University, 3600 University Street, Montreal, QC H3A 2T8, Canada\\
$^{17}$Scuola Normale Superiore, 56126 Pisa, PI, Italy\\
$^{18}$Harvard-Smithsonian Center for Astrophysics, Cambridge, MA\\
$^{19}$Department of Engineering, University of California, Berkeley, CA\\
$^{20}$ARC Centre of Excellence for All-Sky Astrophysics in 3 Dimensions (ASTRO 3D), University of Melbourne, VIC 3010, Australia\\


\bsp	
\label{lastpage}
\end{document}